\documentclass{article}

\usepackage{PRIMEarxiv}

\usepackage[utf8]{inputenc} 
\usepackage{dsfont}

\usepackage{amsfonts}
\usepackage{amsmath}
\usepackage{amssymb}
\usepackage{hyperref}       
\usepackage{url}            
\usepackage{booktabs}       

\usepackage{nicefrac}       
\usepackage{microtype}      
\usepackage{algorithmic}
\usepackage{mathrsfs}
\usepackage{graphicx}
\usepackage{textcomp}
\usepackage{xcolor}
\usepackage{subcaption}
\usepackage{amsthm}

\newtheorem{theorem}{Theorem}

\newtheorem{lemma}{Lemma}

\newtheorem{remark}{Remark}
\newtheorem{assumption*}{Assumption}
\newtheorem{stdassumption*}{Standing Assumption}

\newtheorem{definition}{Definition}
\newtheorem{definition*}{Definition}

\DeclareMathOperator*{\argmax}{arg\,max}

\DeclareMathOperator*{\E}{\mathbb{E}}

\DeclareMathOperator*{\1}{\mathds{1}}

\newcommand{\RS}{\texttt{RS}}
\usepackage{mathtools}
\graphicspath{ {./images/} }

\usepackage[ruled,vlined]{algorithm2e}

\newenvironment{customproof}[1]{\par\noindent\textit{#1:}\quad}{\hfill$\blacksquare$}
\usepackage{cleveref}

\graphicspath{{media/}}     

\title{Preference-Centric Route Recommendation: Equilibrium, Learning, and Provable Efficiency
}

\author{
 Ya-Ting Yang, Yunian Pan, and Quanyan Zhu\\
  ECE Department\\
  New York University \\
  Brooklyn, NY\\
  \texttt{\{yyy4348, yp1170, qz494\}@nyu.edu}
}

\begin{document}
\maketitle

\maketitle

\begin{abstract}

 Traditional approaches to modeling and predicting traffic behavior often rely on Wardrop Equilibrium (WE), assuming non-atomic traffic demand and neglecting correlations in individual decisions. However, the growing role of real-time human feedback and adaptive recommendation systems calls for more expressive equilibrium concepts that better capture user preferences and the stochastic nature of routing behavior.
In this paper, we introduce a preference-centric route recommendation framework grounded in the concept of Borda Coarse Correlated Equilibrium (BCCE), wherein users have no incentive to deviate from recommended strategies when evaluated by Borda scores—pairwise comparisons encoding user preferences. We develop an adaptive algorithm that learns from dueling feedback and show that it achieves $\mathcal{O}(T^{\frac{2}{3}})$ regret, implying convergence to the BCCE under mild assumptions.
We conduct empirical evaluations using a case study to illustrate and justify our theoretical analysis. The results demonstrate the efficacy and practical relevance of our approach.

\end{abstract}

\section{Introduction}
\label{sec:intro}
In today’s transportation networks, the sheer volume of traffic demand poses significant challenges for designing efficient and interpretable route recommendation algorithms on navigational platforms such as Google Maps and Waze. The goal is to optimize user experience by minimizing their “regret,” encouraging them to follow the suggested routes—or, in the language of \textit{congestion games}, to approach a ``\textit{Wardrop Equilibrium}'' (WE) \cite{wardrop1952road}, in which each (non-atomic) traffic demand has no incentive to deviate from its assigned route. This concept provides valuable insight and predictive power for understanding traffic patterns and designing effective recommendation systems (RS).


However, the Wardrop principle has a weak behavioral foundation, as pointed out in several critiques \cite{daganzo1977stochastic,li2024wardropequilibriumboundedlyrational}. It assumes perfect rationality, deterministic route choices, and travel-time-based utilities—assumptions that often fail to capture real-world behavior. In reality, travelers are boundedly rational and influenced by context, with diverse preferences and incentives. For instance, a commuter in New York City might choose the ferry over a faster subway ride on a sunny day. Likewise, riders often prefer certain subway lines over others running on the same tracks due to factors like comfort, crowding, or even aesthetics.

While some of these behaviors can be captured by enriching the cost functions in congestion models (e.g., see \cite{daganzo1977stochastic,di2013boundedly}), any fixed (stochastic) utility function inevitably imposes a (stochastic) partial order over route choices, and assign each user with their most favored route, which might not exist in some scenarios due to the inherent indeterminacy and context sensitivity of user preferences \cite{li2024wardropequilibriumboundedlyrational}.  
Moreover, the increasingly rich user–platform interactions—and the data thereby generated—are often underutilized in traditional models.

Given these limitations, it is necessary to move beyond Wardrop Equilibrium (WE) toward alternative solution concepts. In this work, we introduce an online congestion game framework for recommendation systems that learns from human preference feedback. The learning process covers two steps in each round $t=1, \ldots, T$:

\begin{enumerate} 
\item The recommendation system (RS)—such as Google Maps—samples multiple candidate routes from a recommendation strategy and delivers them to users. \item It then collects user preference feedback on the suggested routes and updates the recommendation distribution for the next round. \end{enumerate}


%
Unlike traditional frameworks that prescribe a single route during the learning process \cite{blum2006,krichene2015convergence}, this framework empowers users with a spectrum of routing choices, capturing more nuanced data-driven interactions between users and the platform.
In this paper, we consider a simplified yet extensible setting in which the recommendation system (RS) suggests a pair of routes in each round and receives binary feedback from users.
The goal is to maximize the likelihood of recommending a \textit{Borda winner} \cite{daunou1803memoire}—a route that is guaranteed to exist. 
We formalize this objective through the notion of \textit{Borda Coarse Correlated Equilibrium} (BCCE)—a joint distribution over user traffic flows and recommendation strategies in which each user, in expectation, is assigned a Borda-winning route and thus has no incentive to deviate.

A key challenge in designing a practical learning scheme under this setup is ensuring that it is scalable, lightweight, provably efficient, and parameter-free \cite{vudong2021}. That is, the gap between the current distribution—over user traffic flows and recommendation strategies—and the target equilibrium should diminish as the number of rounds 
$T$ increases, with a sublinear convergence rate in 
$T$, polynomial dependence on the network size $|\mathcal{S}|$, and low per-round computational cost. Additionally, the algorithm should operate with minimal prior knowledge, such as the distribution of user preferences.

To address this, we propose the Dueling Recommendation (DR), an exponential weights type of algorithm that achieves low user regret of order ($\mathcal{O}(|\mathcal{S}|^{\frac{1}{3}}T^{\frac{2}{3}})$), thus implying the convergence to the BCCE. Notably, classical $(\mathcal{O}(T^{\frac{1}{2}}))$ regret bounds for the universal stochastic mirror descent framework \cite{blum2006, krichene2015convergence,pan2024variational} no longer hold in this setting, as users are occasionally required to explore suboptimal routes to gather information—an instance of bounded rationality embedded in the learning process. We complement our theoretical results with numerical evaluations, aiming to to validate our approach and demonstrate its applicability to realistic, preference-centric route recommendation scenarios.

\section{Related Work}

Our framework builds upon the classical traffic assignment models \cite{leblanc1975efficient}. A substantial body of work has explored the online learning direction \cite{blum2006, krichene2015convergence, vudong2021, pan2023resilience, pan2023stochastic,yang2024adaptive}, where methods—often generalized by mirror descent \cite{krichene2015convergence, pan2024variational}—typically achieve $\mathcal{O}(T^{\frac{1}{2}})$-regret for individual traffic demands. These dynamics naturally lead to convergence toward Nash or Coarse Correlated Equilibria in non-atomic or atomic congestion game settings.

Considerable efforts have extended these learning protocols to accommodate more realistic feedback structures, including bandit and semi-bandit types \cite{panageas2023semi}. A more fundamental concern, however, is the validity of travel-time-utility-based discrete choice models, as questioned in \cite{daganzo1977stochastic, di2013boundedly, li2024wardropequilibriumboundedlyrational}.
Drawing on advances in dueling bandits \cite{saha2021adversarial}, we address this challenge by integrating pairwise comparisons into the congestion game learning framework, enabling a preference-based approach less reliant on cardinal utility assumptions.




\section{Problem Formulation}
\label{sec:problem}

\subsection{Repeated Route Recommendation}

We begin by introducing the key elements of our model: 
    
  {\bf The network and traffic demands:} The urban transportation network can be represented as a graph $\mathcal{G}=(\mathcal{V}, \mathcal{E})$, where $\mathcal{V}$ denotes the set of intersections; $\mathcal{E}$ represents roads between intersections.  Over network $\mathcal{G}$, there are users $u \in \mathcal{U}$ trying to get to destinations $\{ D_u \}_{u \in \mathcal{U}} \subseteq \mathcal{V}$ from origins $\{O_u\}_{u \in \mathcal{U}} \subseteq \mathcal{V}$.
 Each user $u$ then has a set of feasible paths $\mathcal{S}_u$, $\mathcal{S} = \bigcup_{u \in \mathcal{U}} \mathcal{S}_u $ is all the feasible paths.
    
 {\bf Network flow:} Once the users are committed to their path selection profile $ (s_u, s_{-u}) = ( s_u )_{u \in \mathcal{U}}$, a network path\slash edge flow profile $(x, f)$ is formed, 
    \begin{equation*}
    x_p = \sum_{u \in \mathcal{U}} \mathds{1}_{ \{ p = s_u \}}  , p \in \mathcal{S} ; \quad  f_e = \sum_{u \in \mathcal{U}} \mathds{1}_{\{ e \in s_u\}} , e \in \mathcal{E}.
    \end{equation*} 
    In general, $x$ has a linear relation with $f$ depending on how the transportation network is structured. We use $x (s) $ to represent the network flow determined by the path profile since every $x(s)$ corresponds to a unique $f(s)$.
      
     {\bf User preferences:} Each user's preference is modeled by a flow-dependent matrix $\mathscr{P}^{x}_u = [\mathscr{P}^{x}_{u, p,q}]_{p,q \in \mathcal{S}_u }$, where $\mathscr{P}^{x}_{u, p,q} \in [0,1]$ is the probability of user $u$ choosing path $p$ over $q$, given network flow $x$. 
  
     {\bf Recommendation system:} A recommendation system (\RS) \ uses a set of mixed strategies to sample route recommendations for each user. Denote the strategy as $\mathbf{P} := \prod_{u \in \mathcal{U}}\mathbf{P}_u \in 
 \prod_{u \in \mathcal{U}}\Delta (\mathcal{S}_u)$. $\mathbf{P}_u = (p_{u, i})_{i \in \mathcal{S}_u}$ is a probability vector.  The route recommendation cycling periods are discrete $t = 1, \ldots, T \in \mathbb{N}_+$.

Now,  let random variables $i^t_u, j^t_u$ be the two paths sampled from a recommendation strategy $\mathbf{P}^t_u$. We put a superscript $t$ on them to denote the time when they are being sampled, their realizations are denoted as $i,j$.  We abuse the notation a little and denote $\mathscr{P}^{ t}_u := \mathscr{P}^{x^{t-1}}_u. $ 


We adopt the so-called \textit{Borda score} for the quantitative measure of human preferences, which, by assumption, depend on the network flow.
The (shifted) Borda score for user $u$ at time $t$ as: 
\begin{equation}\label{eq:bordascore}
     b^t_u (i) = \frac{1}{|\mathcal{S}_u|}\sum_{ j \neq i, j \in \mathcal{S}_u} \mathscr{P}^{t}_{u, i, j} \quad \quad \forall i \in \mathcal{S}_u
\end{equation}

The user regret is defined as:
\begin{equation}\label{eq:userregret}
  \mathcal{R}_u (T) =  \sum_{t=1}^T b^t_u (i_u^*) - \frac{1}{2} (b^t_u(i^t_u) + b^t_u (j^t_u )),
\end{equation}
where $i^*_u:=  \argmax_{ i \in \mathcal{S}_u } \sum_{t=1}^T b^t_u (i)$ are the paths with the highest Borda score in hindsight. 
The regret for \texttt{RS} is defined as: $
     \mathcal{R}(T) = \sum_{u \in \mathcal{U}} \mathcal{R}_u (T).$

\subsection{Equilibrium Concept and Learning  }
In analogy to the classical definition of Coarse Correlated Equilibrium (CCE), we define Borda Coarse Correlated Equilibrium (BCCE) as a solution concept for our model.
\begin{definition}[Borda Coarse Correlated Equilibrium (BCCE)]
\label{def:bcce}
     Let \( \mathcal{P} \) be a joint distribution over traffic flows and route recommendations, let the Borda scores for each user be defined in \eqref{eq:bordascore}. Then, it is said to be a $\varepsilon$-Borda Coarse Correlated Equilibrium (BCCE) if 
     \begin{equation}\label{eq:bcce}
        \mathbb{E}_{ (x, s) \sim \mathcal{P} } [ b^{x}_u(s_u )  -  b^{x}_u (s_{u}^{\prime}) ] \geq \varepsilon,
     \end{equation}
     for all $u \in \mathcal{U}$ and $ s_u^{\prime} \in \mathcal{S}_u $. When \( \varepsilon = 0 \), we refer to this simply as a BCCE.
\end{definition}

At the heart of this definition is the consistency between the traffic recommendation and the flow formation, given which the users have no incentive to deviate from the recommended routes. Indeed, we can rewrite \eqref{eq:bcce} as  $\mathbb{E}_{ (x,  i, j) \sim \mathcal{P} } [ \frac{1}{2} (b^{x}_u(i) + b^{x}_u(j))  -  b^{x}_u (s_{u}^{\prime}) ] \geq \varepsilon$, meaning that either of the two routes $i, j$ are better than any alternatives $s^{\prime}_u$ with respect to their Borda scores.

\begin{lemma}\label{lem:noregret=bcce}
    Suppose for a sufficiently small $\varepsilon > 0$, there is a sequence of flows and recommendations $\{ x^{t-1}, (i^t_u, j^t_u)_{u \in \mathcal{U}}\}_{t=1}^T$   such that $ \mathcal{R}_u (T) \leq T \varepsilon $ for every user $u \in \mathcal{U}$, then the empirical distribution of flows and route recommendations $\overline{\mathcal{P}}^T$, defined as,
    \begin{equation*}
        \overline{\mathcal{P}}^T (x, i) =  \frac{1}{T}\sum_{t=1}^T \mathds{1}_{\{x^{t-1} = x\}} \frac{ \mathds{1}_{\{ i^t_u = i\}} + \mathds{1}_{\{ j^t_u = i\}}}{2},
    \end{equation*}
    is an $\varepsilon$-BCCE, i.e.,
    \begin{equation}
          \mathbb{E}_{(x, s) \sim \overline{\mathcal{P}}^T }[ b_u^{x} ( s_u )] \geq  \mathbb{E}_{s \sim \overline{\mathbf{P}}^T} [b_u^{x} ( s_u^{\prime} ) ] + \varepsilon, 
    \end{equation}
    for all $u \in \mathcal{U}$ and $s^{\prime}_u \in \mathcal{S}_u$.
\end{lemma}

In designing the algorithm, we assume that the recommender $\texttt{RS}$ is capable of observing the decision-making processes of individual users, i.e., $i^t_u, j^t_u$, and $s^t_u$, and updates the recommendation strategies $\mathbf{P}^t_u$ according to such feedback.

\section{Algorithm and Regret Analysis}

\subsection{The Dueling-Adaptive Recommendation}

\Cref{thealgorithm} is a straightforward adaptation from \cite{saha2021adversarial}, which combines the classical exponential weight strategy of Adversarial bandit learning \cite{auer2002finite}, and the estimates of the Borda score. 
In the dueling recommendation setup, such estimates can only be obtained through binary preference feedback corresponding to a choice from the pair of routes. 
The algorithm estimates the Borda score through \eqref{eq:bordaestimates}, which is an unbiased estimator as shown in \Cref{lem:unbias1}.

\begin{algorithm}
\SetKwInOut{Input}{Input}
\SetAlgoLined
\Input{For every $u \in \mathcal{U}$ initialize: \\
recommendation strategy $\mathbf{P}^1_u \in \Delta(\mathcal{S}_u)$;\\
learning rate $\eta_u \in (0,1)$;\\
exploration rate $\gamma_u \in (0,1)$;}
\For{$t = 1, \ldots, T$}{
  $\textbf{for every $u \in \mathcal{U}$ in parallel}$ \textbf{do:}\\ {
     \ sample two path recommendations $i^t_u, j^t_u \sim \mathbf{P}_u^{t}$ independently with replacement;\\
     \ selects a path $s^t_u =  i^t_u \mathds{1}^t_u + j^t_u (1 - \mathds{1}^t_u)$ where  $$
      {\1}^t_u : = \1\{ i^t_u \succ j^t_u \} \sim \mathrm{Bern} (\mathscr{P}^{t}_{u, i^t_u, j^t_u} ). $$ 
    
     \ The network flow $x^t$ is formed by $(s^t_u)_{u \in \mathcal{U}}$.
  
     \ \texttt{RS} estimates the Borda scores by 
   \begin{equation}\label{eq:bordaestimates}
        \hat{b}^t_u (i) =  \frac{ \1\{ i_u^t = i \} }{|\mathcal{S}_u|p^t_{u,i} } \sum_{ j \in \mathcal{S}_u} \frac{\1\{  j^t_u = j\}}{p^t_{u,j}} {\1}^t_u ;
   \end{equation}
     updates its recommendation strategy 
     \begin{equation*}
       p_{u, i} ^{t+1} =  (1 - \gamma_u) \frac{\exp \left(\eta_u \sum_{\tau = 1}^t  \hat{b}_u^{\tau} (i)\right) }{ \sum_{j \in \mathcal{S}_u} \exp \left(\eta_u \sum_{\tau = 1}^t  \hat{b}_u^{\tau} (j) \right) }  + \frac{\gamma_u }{|\mathcal{S}_u|} .
     \end{equation*}
     }
}
\caption{ {Dueling Recommendation} (DR)} \label{thealgorithm}
\end{algorithm}

At every round $t$, Algorithm \ref{thealgorithm} keeps a recommendation strategy $\mathbf{P}^t$ for every user $u\in \mathcal{U}$, with the weights proportional to the exponentiation of cumulative Borda estimates. Such strategies can be generalized by stochastic mirror descent, see, for example, \cite{vudong2021,krichene2015convergence,pan2024variational}. We also smooth the exponential with an $\gamma_u$-uniform exploration, which constantly encourages \texttt{RS} to assign a constant probability to those ``unfavored'' routes.

\subsection{Provable Efficiency}
\Cref{thm:expectedupperbound} provides a formal guarantee on the efficiency of \cref{thealgorithm}.
\begin{theorem}
\label{thm:expectedupperbound}
Let the \RS\ and each user $u$ follow \Cref{thealgorithm} with $\eta_u = ((\log (|\mathcal{S}_u |))/(T\sqrt{|\mathcal{S}_u |}))^{2/3}$ and $\gamma_u = \sqrt{\eta_u |\mathcal{S}_u |}$.
Then, the expected user regret satisfies:
\begin{equation*}
    \E[\mathcal{R}_u(T)] \le 3(|\mathcal{S}_u |\log |\mathcal{S}_u |)^{1/3}T^{2/3}. 
\end{equation*}
\end{theorem}

\begin{remark}
     Theorem \ref{thm:expectedupperbound} can be straightforwardly strengthened to a high‑probability bound, which in turn implies a $\mathcal{O}(\max_{u \in \mathcal{U}}(|\mathcal{S}_u |\log |\mathcal{S}_u|)^{\frac{1}{3}} T^{-\frac{1}{3}})$ convergence to the BCCE. Due to space constraints, we defer a formal statement and proof of the high-probability bound to future work.
\end{remark}

The proof of \Cref{thm:expectedupperbound} relies on the quantitative analysis of the Borda scores estimated by $\texttt{RS}$, including their magnitude, unbiasedness, and bounded variances. 

\Cref{lem:boundedmag} demonstrates that, as a result of the $\gamma_u$-uniform exploration mechanism, the magnitude of  $\tilde{b}^t_u $ scales proportionally with the cardinality of $|\mathcal{S}_u|$.

\begin{lemma}[Bounded Magnitude of Borda Estimator] \label{lem:boundedmag}
    In \Cref{thealgorithm} with uniform exploration rate \( \gamma_u \), for any \( t \), the Borda estimator \( \hat{b}^t_u(i) \) for user \( u \) and route \( i \in S_u \) satisfies:
     \begin{equation}
          \tilde{b}^t_u  (i) \leq \frac{|\mathcal{S}_u|}{\gamma_u^2}.
     \end{equation}
\end{lemma}

Let $ \mathcal{F}_{t-1} = \left( x^0, \mathbf{P}^1, (i^1_u, j^t_u)_{u \in \mathcal{U}}, \mathds{1}^t_u, \ldots, x^{t-1}, \mathbf{P}^t \right)$ be the filtration up to time $t-1$ for the formalism. we show that \eqref{eq:bordaestimates} is an unbiased estimate, as in \Cref{lem:unbias1}.

\begin{lemma}[Unbiasedness] \label{lem:unbias1}
  The Borda estimator \( \hat{b}^t_u(i) \), computed via pairwise feedback at time \( t \), is unbiased:
\[
\mathbb{E}[\hat{b}^t_u(i) \mid \mathcal{F}_{t-1}] = b^t_u(i) .
\]
\end{lemma}

Through \Cref{lem:unbias2}, we are able to express the instantaneous regret at time $t$ as a function of the Borda score. 
\begin{lemma}[Expected Borda score]\label{lem:unbias2}
  Given the recommendation strategy $\mathbf{P}^t_u$ at time $t$, we have the following holds
 \begin{equation}
     \mathbb{E}_{\mathcal{F}_{t}} \left[ \langle \mathbf{P}^t_{u} ,\tilde{b}^{t}_{u} \rangle \right] = \mathbb{E}_{\mathcal{F}_{t-1}}  \left[ \mathbb{E}_{s_u \sim \mathbf{P}^t_u}  b^{t}_u (s_u)  \big \vert \mathcal{F}_{t-1} \right],
 \end{equation}  
 for any given time $t = 1, \ldots, T$.
\end{lemma}

Another critical step is to show that the estimated Borda scores have finite second moments, as demonstrated in in \Cref{lem:finitesecondmoment}.

\begin{lemma}[Bounded second moment]\label{lem:finitesecondmoment}
Given the recommendation strategy $\mathbf{P}^t_u$ at time $t$, we have the following holds 
  \begin{equation}\label{eq:finitesecondmoment}
      \mathbb{E}[  \sum_{  i \in \mathcal{S}_u } p^t_{u,i} (\tilde{b}^{t}_u(i))^2 ] \leq \frac{|\mathcal{S}_u |}{\gamma_u}, 
  \end{equation}
  for any given time $t= 1, \ldots, T$. 

\end{lemma}

{\bf Proof overview for \Cref{thm:expectedupperbound}:}
Equipped with the machinery above, we hereby give a brief summary of the proof. We first make use of the fact that $$
\begin{aligned}
    \mathbb{E}_{\mathcal{F}_{t}} \left[ b^t_u({i^t_u }) + b^t_u({j^t_u })\right] & = \mathbb{E}_{\mathcal{F}_{t-1}} \left[ \mathbb{E} [ b^t_u({i^t_u }) + b^t_u({j^t_u }) ] \big| \mathcal{F}_{t-1}\right] \\ 
    & =\mathbb{E}_{\mathcal{F}_{t-1}} \left[ \mathbb{E}_{s_u \sim \mathbf{P}^t} [ b^t_u({s_u })  ] \big| \mathcal{F}_{t-1}\right] ,
\end{aligned}$$
to rewrite the regret in the form,  
    \begin{align*}
    \mathbb{E}_{\mathcal{F}_{T}} \left[ \mathcal{R}_u(T) \right] & =  \sum_{t=1}^T \mathbb{E}_{\mathcal{F}_{t-1}} \left[  
  b^t_u (i^*_u)  - \mathbb{E}_{s_u \sim \mathbf{P}^t} [ b^t_u({s_u })  ]   \big|  \mathcal{F}_{t-1} \right] .
    \end{align*}
We then apply the standard adversarial‑bandit analysis, which bounds the total regret in terms of the cumulative instantaneous variance introduced by our unbiased loss estimates at each update. The crucial step is to bound
$    (1 - \gamma_u)\sum_{t=1}^T \tilde{b}^t_u (i) - \sum_{t=1}^T (\mathbf{P}^t)^{\top} \tilde{b}^t_u$ using the sum $\frac{\log |\mathcal{S}_u|}{ \eta_u} +  \eta_u \sum_{t=1}^T \sum_{i \in \mathcal{S}_u }   \tilde{p}^t_{u,i} (\tilde{b}^t_u(i))^2.$
The proof then is completed by optimizing $\eta_u $ and $\gamma_u$.

\section{Case Study}
\label{sec:casestudy}

\begin{figure}[htbp]
    \centering
    \includegraphics[width=\linewidth]{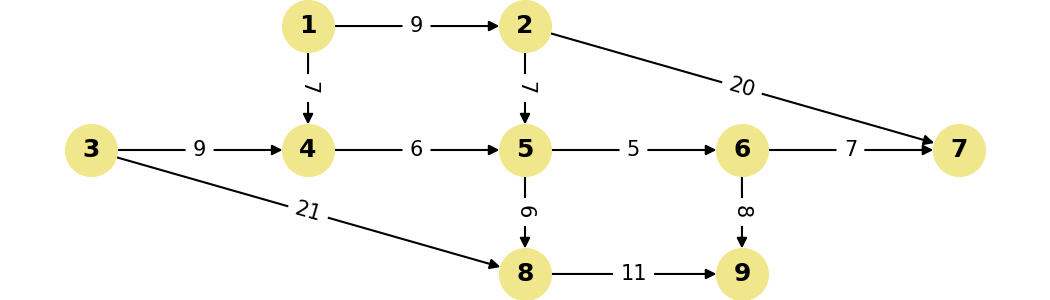}
    \caption{In this example network, there are three O-D pairs (1-7), (1-9), and (3-9). The edge values represent the free-flow travel time of the corresponding road segment.}
    \label{fig:network}
\end{figure}

We illustrate \Cref{thealgorithm} through a case study where the network is illustrated in \Cref{fig:network}. Intuitively, the classic selfish-routing task would require balancing the network flows between center paths (such as (1-4-5-6-7) and (1-4-5-6-9)) and side paths ((3-8-9) and (1-2-7)).

We consider an example scenario where the users' preference matrices can be calculated based on the travel time costs. Specifically, for user $u$ and path $i, j \in \mathcal{S}_u$, the matrix entry $$\mathscr{P}^t_{u, i, j} := \frac{c_{u, j}(x^{t-1})}{ c_{u,i}(x^{t-1}) + c_{u,j} (x^{t-1})}, $$ where $c_{u,i} (x^{t-1})$ and $c_{u,j}(x^{t-1})$ represent the perceived travel costs for user $u$ on paths $i$ and $j$, respectively, given the last round network path flow $x^{t-1}$, which, in our experiment, is set to be of the linear form $c_{u, i}(x)=a_u\sum_{e \in i}c_e(f_e) + b_u$, where $a_u, b_u$ represent the user's sensitivity to travel time costs, and $f$ is the edge flow corresponding to path flow $x$. We adopt the standard Bureau of Public Roads (BPR) function as $c_e(\cdot)$, 
\begin{equation*}
    c_e(f_e):=t_e\left(1+\alpha\left(\frac{x_e}{k_e}\right)^\beta\right),
\end{equation*}
where $t_e$ are the free travel time, $k_e \in \mathbb{R}_{+}$ are the flow capacity of edge $e$, parameters $\alpha, \beta \in \mathbb{R}_{\ge 0}$ are set to be $0.35$ and $2$, respectively.

\begin{figure}
    \centering
    \includegraphics[width=.9\linewidth]{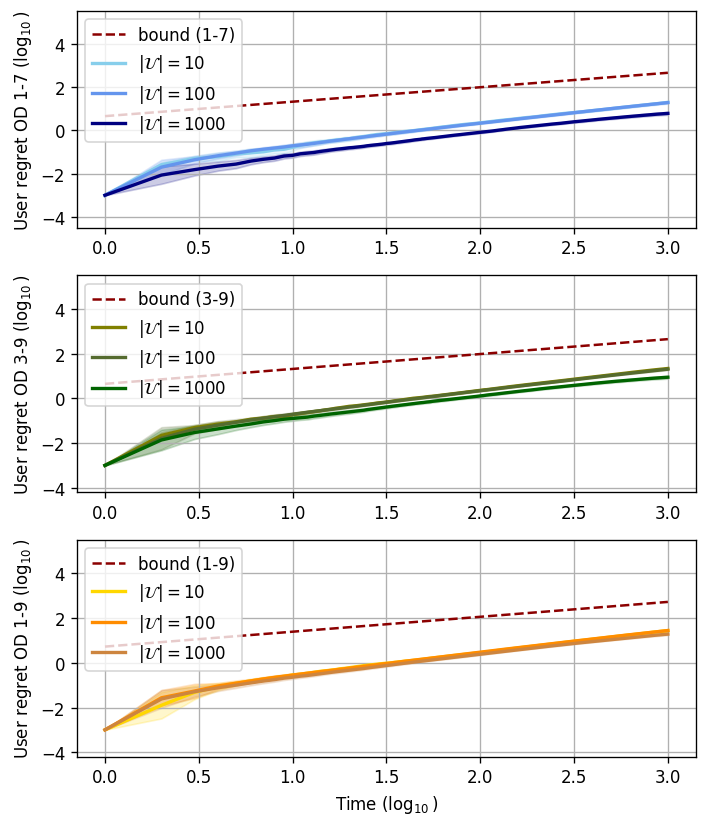}
    \caption{The regret plot of different user population configurations, where we plot the mean regret with standard deviations for different OD populations.}
    \label{fig:regretplot}
\end{figure}

We compare three user population configurations:  $| \mathcal{U}| = 10, 100, 1000$. In each case, user demand is proportionally distributed across three OD pairs: 20\% to (1–7), 30\% to (1–9), and 50\% to (3–9).
The road capacity is set to be fixed at $500$. The user learning parameters $\eta_u$ and $\gamma_u$ are set to align with \Cref{thm:expectedupperbound}.  
The corresponding log-log regret plot for each OD population is shown in \Cref{fig:regretplot}. 

In our experiments, we observed that across different user configurations, the user regret consistently follows a trend that gradually aligns with the theoretical upper bound, supporting our analytical results.
We also found that as the number of users increases, the average Borda regret tends to slightly decrease. 
A possible explanation is that with more users, the user preferences—shaped by travel time—becomes more significant with respect to traffic volume, which in turn amplifies the differences in Borda scores. This heightened interaction appears to accelerate convergence toward the BCCE.

\section{Conclusion and Future Work}

In this work, we have proposed a novel preference-centric learning framework for route recommendation. We formulated BCCE as a novel equilibrium concept and developed a dueling-feedback algorithm with sublinear regret guarantees to approach BCCE. 
We provide both theoretical and experimental justifications for our methods.
Our approach captures richer, behaviorally grounded user–platform interactions beyond classical utility-based models.

Future work includes extending the framework to subset-wise recommendations with partially observable ranking feedback, establishing performance guarantees, and exploring theoretical connections to the non-atomic setting.

\bibliographystyle{abbrv}
\bibliography{ref}

\appendix

\section{Proofs}

\begin{customproof}{Proofs of Lemma \ref{lem:noregret=bcce}}

Fixing a sequence of flows and recommendations, $x^{t-1}, (i^t_u, j^t_u)_{u \in \mathcal{U}}$, let $i^*_u$ be the hindsight Borda winner for user $u$, $i^*_u$ is fixed given the network flow and route recommendation sequence. 

The no-regret dynamics says, for any user $u \in \mathcal{U}$, we have, 
\begin{equation*}
    \begin{aligned}
   & \quad  \mathcal{R}_T   = \frac{1}{T} \sum_{t=0}^T   b^{t}_u (i_u^*) - \frac{1}{2} (b^{t}_u(i^t_u) + b^{t}_u (j^t_u )) 
     \\ 
     & =  \frac{1}{T} \sum_{t=0}^T \left( b^t_u (i_u^*) - \sum_{i \in \mathcal{S}_u }b^t_u(i) \frac{\mathds{1}_{\{i^t_u = i\}} + \mathds{1}_{\{j^t_u = i\}} }{2} \right)  \\
     & =  \sum_{x}\left[\frac{1}{T} \sum_{t=0}^T\mathbb{E}_{ (s_u) \sim \overline{\mathbf{P}}^t_u} \left( b^{x}_u(i_u^*) - b^{x}_u(s_u)\right) \mathds{1}_{\{x = s^{t-1}\}}\right]   \\
       & =  \mathbb{E}_{ (x, (s_u)_{u\in\mathcal{U}} ) \sim \overline{\mathcal{P}}^T_u} \left[ b^{x}_u(i_u^*) - b^{x}_u(s_u)  \right] \leq   \varepsilon.
    \end{aligned}
\end{equation*}
Therefore, by \Cref{def:bcce}, the empirical distribution $\overline{\mathcal{P}}^T$ is a $\varepsilon$-BCCE.

\end{customproof}

\begin{customproof}{Proof of \Cref{lem:boundedmag}}
  Following the definition of $ \tilde{b}^{t}_u$, we have that, for every $u \in \mathcal{U}$, since $p^t_{u,i} \geq \frac{\gamma}{|\mathcal{S}_u|}$ for every $i$. 
     \begin{align*}
          \tilde{b}^{t}_u (i) & =  \frac{ \1\{ i_u^t = i \} }{|\mathcal{S}_u|p^t_{u,i} } \sum_{ j \in \mathcal{S}_u} \frac{\1\{  j^t_u = j\}}{p^t_{u,j}} {\1}^t_u \\ 
          &\leq \frac{1}{ |\mathcal{S}_u| \frac{\gamma}{|\mathcal{S}_u |}}  \sum_{ j  \in \mathcal{S}_u }  \frac{\mathds{1}_{\{ j^t_u  = j \}}}{\frac{\gamma}{|\mathcal{S}_u |}}
          \\ 
          &  \leq  \frac{ 1}{ \gamma} \sum_{j \in \mathcal{S}_u } \frac{|\mathcal{S}_u |}{\gamma } \mathds{1}_{\{ j^t_u  = j \}} leq \frac{|\mathcal{S}_u|}{\gamma^2 } , 
     \end{align*}
     which concludes the proof.
\end{customproof}

\begin{customproof}{Proof of \Cref{lem:unbias1}}
  For every $ u \in \mathcal{U}$, $i \in \mathcal{S}_u$,  we have
    \begin{align*}
       & \quad  \mathbb{E}[ \tilde{b}^{t}_u (i)]  = \mathbb{E}_{\mathcal{F}_t} \left[ \frac{  \mathds{1}_{ \{ i^t_u  = i \} }}{|\mathcal{S}_u|  p^t_{u,i} } \sum_{ j \in \mathcal{S}_u} \frac{\mathds{1}_{\{ j^t_u = j\}}  }{p^t_{u,j} } \mathds{1}^t_u \right]  
        \\
        & =  \mathbb{E}_{\mathcal{F}_{t-1} } \left[ \mathbb{E}_{i^t_u, j^t_u }\sum_{j \in \mathcal{S}_u} \frac{ \mathds{1}_{\{ i^t_u = i \}} \mathds{1}_{ \{ i^t_u \succ j^t_u  \} } \mathds{1}_{\{ j^t_u = j \}}}{ |\mathcal{S}_u| p^t_{u,i} p^t_{u,j}} | \mathcal{F}_{t-1} \right] 
        \\
        & =  \mathbb{E}_{\mathcal{F}_{t-1} } \left[   \frac{ \mathbb{E}\mathds{1}_{ \{ i^t_u = i\} }}{|\mathcal{S}_u|p^t_{u,i}} \sum_{ j \in \mathcal{S}_u }   \frac{ \mathbb{E}\mathds{1}_{\{ j^t_u = j , i \succ j \}  } }{p^t_{u,j}}    | \mathcal{F}_{t-1}\right]
        \\
        & =  \frac{1}{|\mathcal{S}_u|}\mathbb{E}_{\mathcal{F}_{t-1} } \left( \frac{p^t_{u,i}}{p^t_{u,i}}\sum_{ j \neq i, j \in \mathcal{S}_u } p^t_{u,j} \frac{ \mathbb{E}[\mathds{1}_{\{ i \succ j \}  } ] }{p^t_{u,j}}  \right) 
        \\
        & =\frac{1}{ | \mathcal{S}_u|}  \mathbb{E}_{\mathcal{F}_{t-1} } \left[\sum_{ j \neq i,  j \in \mathcal{S}_u } \mathscr{P}^{t}_{u, i,j}  \right]
        =  \mathbb{E}_{\mathcal{F}_{t-1} } [b^t_u] , 
    \end{align*}
    where the third last inequality comes from that if $ i \succ j$ it must be that $j \neq i$.  This concludes that $\mathbb{E} [\tilde{b}^t_u] = b^t_u $ since it holds for every $i \in \mathcal{S}_u $.
\end{customproof}

\begin{customproof}{Proof of \Cref{lem:unbias2}}
     Following the similar reasoning from the proof of \Cref{lem:unbias1}, we have:
     \begin{align*}
      & \quad   \mathbb{E}_{\mathcal{F}_{t}} \left[ \langle \mathbf{P}^t_{u} ,\tilde{b}^{t}_{u} \rangle \right]  =  \mathbb{E}_{\mathcal{F}_{t}} \left[ \sum_{i \in \mathcal{S}_u }p^t_{u,i} \tilde{b}^{t}_{u}(i) \right] \\
    & =  \mathbb{E}_{\mathcal{F}_{t-1}} \left[ \sum_{i \in \mathcal{S}_u }p^t_{u,i} \mathbb{E}_{ i^t_u, j^t_u, s^t_u} \left[ \tilde{b}^{t}_{u}(i) \big \vert \mathcal{F}_{t-1} \right]\right] 
    \\
    & =  \mathbb{E}_{\mathcal{F}_{t-1}} \left[ \sum_{i \in \mathcal{S}_u }p^t_{u,i} \mathbb{E}_{i^t_u, j^t_u } \left[ \sum_{j \in \mathcal{S}_u} \frac{  \mathds{1}_{ \{ i^t_u \succ j^t_u  \} } }{ |\mathcal{S}_u| p^t_{u,i} p^t_{u,j}}  \big \vert \mathcal{F}_{t-1} \right]\right]  
\\ 
& =  \mathbb{E}_{\mathcal{F}_{t-1}} \left[ \sum_{i \in \mathcal{S}_u }p^t_{u,i}   \left[   \frac{ \mathbb{E}\mathds{1}_{ \{ i^t_u = i\} }}{|\mathcal{S}_u|p^t_{u,i}} \sum_{ j \in \mathcal{S}_u }   \frac{ \mathbb{E}\mathds{1}_{\{ j^t_u = j , i \succ j \}  } }{p^t_{u,j}}    \big| \mathcal{F}_{t-1}\right]\right] 
\\ 
& = \mathbb{E}_{\mathcal{F}_{t-1}} \left[ \sum_{i \in \mathcal{S}_u }p^t_{u,i}   b^t_u(i) \big| \mathcal{F}_{t-1}  \right] \\
& = \mathbb{E}_{\mathcal{F}_{t-1}} \bigg[  \mathbb{E}_{ s_u \sim \mathbf{P}^t} \left[b^t_u(s_u) \right] \big| \mathcal{F}_{t-1} \bigg], 
     \end{align*}
which concludes the proof. 
\end{customproof}

\begin{customproof}{Proof of \Cref{lem:finitesecondmoment}}
We first recall that:
    $ \tilde{b}^{t}_u (i):= \frac{\mathds{1}_{\{i^t_u  =  i\}} }{|\mathcal{S}_u| p^t_{u,i}} \sum_{j \in \mathcal{S}_u} \frac{\mathds{1}_{\{ j^t_u = j \}}}{p^t_{u,j} }  \mathds{1}^t_u $, 
    as a substitute for $\tilde{b}$ in expression \eqref{eq:finitesecondmoment}. 
Then, we have
    \begin{align*}
    & \quad  \mathbb{E}[  \sum_{  i \in \mathcal{S}_u } p^t_{u,i} (\tilde{b}^{t}_u(i))^2 ]  \\ 
     &  =   \mathbb{E}_{\mathcal{F}_{t-1}}\left[   \sum_{i  \in \mathcal{S}_u} p^t_{u,i} 
 \mathbb{E}_{} \left[ \sum_{j \in \mathcal{S}_u } \frac{\mathds{1}_{\{i^t_u  =  i\}} \mathds{1}^t_u \mathds{1}_{\{ j^t_u = j \}}}{ |\mathcal{S}_u| p^t_{u,i} p^t_{u,j}} | \mathcal{F}_{t-1}\right]^2\right] \\
 & = \mathbb{E}_{\mathcal{F}_{t-1}} \Bigg[   \sum_{i  \in \mathcal{S}_u} p^t_{u,i}   \sum_{j \in \mathcal{S}_u } 
 \mathbb{E}_{i^t_u , j^t_u} \left[ \frac{\mathds{1}_{ \{ i^t_u  = i\} }\mathds{1}_{\{ j^t_u = j \}}}{ |\mathcal{S}_u |^2(p^t_{u,i})^2 (p^t_{u,j})^2} \right] \\ & \quad \quad 
 \mathbb{E} \left[ (\mathds{1}^t_u)^2  \big\vert i^t_u, j^t_u  \right] | \mathcal{F}_{t-1}\Big] \Bigg] \\
 &  \leq  \frac{1}{|\mathcal{S}_u|^2 }\mathbb{E}_{\mathcal{F}_{t-1}}   \left[ \sum_{ i \in \mathcal{S}_u }\sum_{j \in \mathcal{S}_u } 
 \frac{1}{ p^t_{u,j} } \bigg\vert \mathcal{F}_{t-1}\right] \\
 & \leq   \frac{1}{|\mathcal{S}_u |^2 }\sum_{ i  \in \mathcal{S}_u}\frac{|\mathcal{S}_u |^2 }{\gamma}= \frac{| \mathcal{S}_u |}{\gamma} .
    \end{align*}
    The argument follows similar to the proof of \Cref{lem:unbias2}. 
\end{customproof}

\begin{customproof}{Proof of \Cref{thm:expectedupperbound}}
We define random variables for all users $u \in \mathcal{U}$, 
\begin{align*}
    Z^t_u = \sum_{ i  \in \mathcal{S}_u} \exp(\eta_u \sum_{\tau =1}^t \tilde{b}^{\tau}_u (i)), \quad t = 1, \ldots, T \text{ and } Z^0_u = |\mathcal{S}_u|.
\end{align*}
We have, for an arbitrary $i \in \mathcal{S}_u$, $
    \exp( \eta_u  \sum_{\tau=1}^t \tilde{b}^{\tau}_u (i))  \leq  \sum_{ i \in \mathcal{S}_u}  \exp( \eta_u \sum_{\tau=1}^t \tilde{b}^{\tau}_u (i)) = Z^t_u = Z^0_u \prod_{\tau=1}^t \frac{Z^{\tau}_u}{Z^{\tau-1}_u } .$
Now the ratio of the product can be written in terms of $\tilde{p}^t_{u,i} : = \cfrac{\exp(\eta_u \sum_{\tau=1}^{t-1} \tilde{b}^\tau_u (i))}{\sum_{j \in \mathcal{S}_u }\exp(\eta_u \sum_{\tau=1}^{t-1}\tilde{b}^\tau_u (i))}$, 
we have $
    \frac{Z^t_u }{Z^{t-1}_u }  = \sum_{i \in \mathcal{S}_u} \frac{\exp(\eta_u \sum_{\tau=1}^{t-1} \tilde{b}^{\tau}_u (i) )}{Z^{t-1}_u }  \exp(\eta_u \tilde{b}^t_u (i) )  = \sum_{ i \in \mathcal{S}_u } \tilde{p}^t_{u,i}  \exp(\eta_u \tilde{b}^t_u (i) ). $
Using the fact that $\exp(x) \leq 1 +x+ x^2 $ for all $x \leq 1$ and $1 + x \leq \exp(x)$ for all $ x \in \mathbb{R}$, we have, since $\tilde{b}^t_u (i)\leq 1$ for any $i \in \mathcal{S}_u$,
\begin{align*}
    \frac{Z^t_u }{Z^{t-1}_u} & \leq 1 + \eta_u \sum_{i \in \mathcal{S}_u } \tilde{p}^t_{u,i} \tilde{b}^t_u (i) + \eta_u^2 \sum_{ i \in \mathcal{S}_u} \tilde{p}^t_{u,i} (\tilde{b}^t_u(i))^2  \\
     & \leq \exp\left(  \eta_u \sum_{i \in \mathcal{S}_u } \tilde{p}^t_{u,i} \tilde{b}^t_u (i) + \eta_u^2 \sum_{ i \in \mathcal{S}_u} \tilde{p}^t_{u,i} (\tilde{b}^t_u(i))^2   \right).
\end{align*}
Therefore, 
\begin{align*}
    & \quad \exp\left(\eta_u \sum_{t=1}^T \tilde{b}^{\tau}_u(i^*_u) \right)
    \leq \sum_{ i \in \mathcal{S}_u}  \exp( \eta_u \sum_{t=1}^T \tilde{b}^{t}_u (i)) \\
 & \leq |\mathcal{S}_u| \prod_{t=1}^T \exp\left(  \eta_u \sum_{i \in \mathcal{S}_u } \tilde{p}^t_{u,i} \tilde{b}^t_u (i) + \eta_u^2 \sum_{ i \in \mathcal{S}_u} \tilde{p}^t_{u,i} (\tilde{b}^t_u(i))^2    \right) \\
   &  = |\mathcal{S}_u| \exp\left(  \eta_u \sum_{t=1}^T \sum_{i \in \mathcal{S}_u } \left(\tilde{p}^t_{u,i} \tilde{b}^t_u (i) + \eta_u^2  \tilde{p}^t_{u,i} (\tilde{b}^t_u(i))^2 \right)   \right),  
\end{align*}    
which leads to 
\begin{align*}
    \sum_{t=1}^T \tilde{b}^t_u (i) \leq \frac{\log |\mathcal{S}_u|}{ \eta_u} + \sum_{t=1}^T \sum_{i \in \mathcal{S}_u } \left(\tilde{p}^t_{u,i} \tilde{b}^t_u (i) + \eta_u  \tilde{p}^t_{u,i} (\tilde{b}^t_u(i))^2 \right). 
\end{align*}
Hence, we arrive at that, starting from flow $x^0$, for a fixed sequence of dueling recommendations $\{(i^t_u, j^t_u )_{u \in \mathcal{U}}\}_{t=1}^T$ and routing profiles $\{(s^t_u)_{u \in \mathcal{U}}\}_{t=1}^{T-1}$, 
\begin{align*}
    & \quad  \sum_{t=1}^T \tilde{b}^t_u (i) - \sum_{t=1}^T \sum_{i \in \mathcal{S}_u} \tilde{p}^t_{u,i} \tilde{b}^t_u(i) \\
    & \leq  \frac{\log |\mathcal{S}_u|}{ \eta_u} +  \eta_u \sum_{t=1}^T \sum_{i \in \mathcal{S}_u }   \tilde{p}^t_{u,i} (\tilde{b}^t_u(i))^2 . 
\end{align*}
   Now, using the fact that $\tilde{p}^t_{u,i} = (p^t_{u,i} - \frac{\gamma_u}{|\mathcal{S}_u|}) / (1 - \gamma_u)$, multiplying the left side with $1 - \gamma_u < 1$ and subtract the positive term $\frac{\gamma_u}{|\mathcal{S}_u |}\sum_{t=1}^T\sum_{i \in \mathcal{S}_u}  \tilde{b}^t_u(i) $, we obtain,
   \begin{align*}
     & \quad  (1 - \gamma_u)\sum_{t=1}^T \tilde{b}^t_u (i) - \sum_{t=1}^T (\mathbf{P}^t)^{\top} \tilde{b}^t_u \\
    & \leq  \frac{\log |\mathcal{S}_u|}{ \eta_u} +  \eta_u \sum_{t=1}^T \sum_{i \in \mathcal{S}_u }   \tilde{p}^t_{u,i} (\tilde{b}^t_u(i))^2.
   \end{align*}

Taking expectation on both sides we have,
\begin{align*}
    & \quad  (1 - \gamma_u)\sum_{t=1}^T \mathbb{E}_{\mathcal{F}_T}\left[\tilde{b}^t_u (i) \right] - \sum_{t=1}^T \mathbb{E}_{\mathcal{F}_T}\left[(\mathbf{P}^t)^{\top} \tilde{b}^t_u \right]\\
    & \leq \frac{\log |\mathcal{S}_u|}{ \eta_u} +  \mathbb{E}_{\mathcal{F}_T}\left[ \eta_u \sum_{t=1}^T \sum_{i \in \mathcal{S}_u }   \tilde{p}^t_{u,i} (\tilde{b}^t_u(i))^2 \right].
\end{align*}

Recall \Cref{lem:unbias1}, substitute $i$ with $i^*_u = \argmax_{i \in \mathcal{S}_u} b^t_u (i)$, and then recall\Cref{lem:unbias2} we get:
\begin{align*}
   & \quad  (1 - \gamma_u)\sum_{t=1}^T \mathbb{E}_{\mathcal{F}_{t-1}}\left[{b}^t_u (i^*_u) \right] - \sum_{t=1}^T \mathbb{E}_{\mathcal{F}_{t-1}}\left[(\mathbf{P}^t)^{\top} \tilde{b}^t_u  \big\vert \mathcal{F}_{t-1}\right]\\
    & \leq \frac{\log |\mathcal{S}_u|}{ \eta_u} +  \mathbb{E}_{\mathcal{F}_t}\left[ \eta_u \sum_{t=1}^T \sum_{i \in \mathcal{S}_u }   \tilde{p}^t_{u,i} (\tilde{b}^t_u(i))^2 \right] \\
    & \leq \frac{\log |\mathcal{S}_u|}{ \eta_u} + \eta_u \sum_{t=1}^T \frac{|\mathcal{S}_u|}{\gamma_u },
\end{align*}
which leads to, along with the fact that $b^t_u (i) \leq 1$, 
\begin{align*}
     \mathbb{E}_{\mathcal{F}_T} [ \mathcal{R}_u(T)] \leq \gamma_u T + \frac{\log|\mathcal{S}_u|}{ \gamma_u } + \eta_u T \frac{|\mathcal{S}_u|}{\gamma_u} , 
\end{align*}
Choosing $\gamma_u = {\sqrt{\eta_u |\mathcal{S}_u|}}$ and $\eta_u = \left(\frac{\log|\mathcal{S}_u|}{ T \sqrt{|\mathcal{S}_u|}}\right)^{\frac{2}{3}} \in (0,1)$, with $ |\mathcal{S}_u| \log |\mathcal{S}_u | \leq T$, $\gamma_u \in (0,1)$ as well. Combining the terms together we conclude that, 
\begin{align*}
    \mathbb{E} [ \mathcal{R}_u(T)] \leq  3 T^{\frac{2}{3}} (|\mathcal{S}_u| \log |\mathcal{S}_u|)^{\frac{1}{3}}.
\end{align*}
\end{customproof}

\end{document}